\documentclass[twocolumn,superscriptaddress,floatfix]{revtex4}
\usepackage{amssymb,amsfonts,amsmath}
\usepackage{color}
\usepackage{graphicx}

\begin{document}

\title{A measure of individual role in collective dynamics}

\author{Konstantin Klemm}
\affiliation{Bioinformatics, Institute for Computer Science,
Leipzig University, H\"{a}rtelstrasse 16-18, 04107 Leipzig, Germany}
\affiliation{Instituto de F\'isica Interdisciplinar y Sistemas Complejos IFISC
(CSIC-UIB), Campus Universitat Illes Balears, E07122 Palma de Mallorca, Spain}

\author{M.\ \'{A}ngeles Serrano}
\affiliation{Departament de Qu\'imica F\'isica, Universitat de Barcelona,
Mart\'i i Franqu\`es 1, 08028 Barcelona, Spain},

\author{V\'\i ctor M. Egu\'\i luz}
\affiliation{Instituto de F\'isica Interdisciplinar y Sistemas Complejos IFISC
(CSIC-UIB), Campus Universitat Illes Balears, E07122 Palma de Mallorca, Spain}

\author{Maxi San Miguel}
\affiliation{Instituto de F\'isica Interdisciplinar y Sistemas Complejos IFISC
(CSIC-UIB), Campus Universitat Illes Balears, E07122 Palma de Mallorca, Spain}

\date{\today}

\begin{abstract}
Identifying key players in collective dynamics remains a
challenge in several research fields, from
the efficient dissemination of ideas to drug target discovery
in biomedical problems. The difficulty lies at several levels:
how to single out the role of individual elements in such
intermingled systems, or which is the best way to quantify
their importance. Centrality measures describe a node's
importance by its position in a network. The key
issue obviated is that the contribution of a node to the
collective behavior is not uniquely determined by the
structure of the system but it is a result of the interplay
between dynamics and network structure. We show that {\em
dynamical influence} measures explicitly  how strongly a
node's dynamical state affects collective behavior. For
critical spreading, dynamical influence targets nodes
according to their spreading capabilities. For
diffusive processes it quantifies how efficiently real
systems may be controlled by manipulating a single node.
\end{abstract}

\maketitle

Complex networks are a groundbreaking concept that is helping to understand the
behavior of many chemical, biological, social and technical systems
\cite{Dorogovtsev:2003,Albert:2002}.
Network representations are particularly suitable for systems where
heterogeneity dominates and are crucial for dynamics \cite{Pastor-Satorras:2001},
where a few nodes are usually considered as the most
important. Oftentimes, node importance is correlated with centrality measures,
local \cite{Barabasi:1999,Jeong:2001} or global \cite{Wuchty:2003}, which usually do not 
explicitly account for the dynamics as they are generally based on a purely topological perspective.
However, dynamics is fundamental in assessing the impact of individual elements in global
performance and in controllability problems \cite{Liu:2011}. Here, we show that
{\em dynamical influence} is a centrality measure able to
quantify how strongly a node's dynamical state affects the collective behavior of
a system, taking explicitly into account the interplay between structure and
dynamics in complex networks. We prove that it applies equally well to a variety of
families of dynamical models, from spreading phenomena at the critical point
to diffusive processes and and continuous-time dynamical system such as the Kuramoto model
and the Roessler chaotic dynamics.

Classical centrality measures in complex networks --like the degree or number of
neighbors a node interacts with \cite{Barabasi:1999,Jeong:2001}, betweenness
centrality \cite{Freeman:1977} counting the number of shortest paths through a
certain node, eigenvector centrality \cite{Bonacich:1972} based on the idea that
relations with more influential neighbors confer greater importance, or the
$k$-shell decomposition \cite{Bollobas:1984} that correlates with the outcome of
supercritical spreading originating in specific nodes \cite{Seidman:1983,Dorogovtsev:2006,Kitsak:2010}-- rely
only on topology, even if an underlying process can be indirectly associated in
some cases. In contrast, the impact of  individual elements in the global
performance of the system inevitably depends on the specificities of the
dynamics. Targeting individuals for vaccination strategies in epidemic processes
is not the same as selecting electrical stimulation sites in the brain in order
to suppress epileptic seizures. In this respect, a Laplacian-based centrality
measure \cite{Masuda:2009b,Masuda:2010,Golub:2010}, closely related to
PageRank \cite{Brin:1998}, has been proposed recently to assess the importance
of complex network nodes in specific dynamical models.

In this work, we provide a  general and rigorous framework where
dynamical influence is defined as a centrality measure both on directed and on
undirected complex networks and applies to a variety of families of dynamical
models, including epidemic spreading models like the
susceptible-infected-removed (SIR), the susceptible-infected-susceptible (SIS), and
the contact process, the Ising model, and diffusive processes like the voter
model or phase coupled oscillators. In all cases, dynamical influence is
calculated as the leading left eigenvector of a characteristic matrix
that encodes the interplay between topology and dynamics.

\section*{\large Results}


\subsection*{Defining dynamical influence}

We focus on systems of $N$ time-dependent real variables $x=(x_1,\dots,x_N)$
with coupled linear dynamics specified by a $N \times N$ real matrix $M$
\begin{equation} \label{eq:main}
\dot{x} = Mx.
\end{equation}
A first classification of the dynamics is obtained by considering the largest
eigenvalue $\mu_{\text max}$ of $M$. For $\mu_{\text max}<0$, $x(t)$ converges
to a null vector that represents a stable fixed point solution; for $\mu_{\text
max}>0$, indefinite growth from almost all initial conditions is observed.
Suppose that $M$ is such that a non-degenerate $\mu_{\text max}=0$ exists.
Then,
the scalar product $\phi_c=c \cdot x$
is a conserved quantity, where $c$ is the left eigenvector of $M$ for
$\mu_{max}$,
\begin{equation} \label{eq:asymptotic}
\frac{\text{d}\phi_c}{\text{d}t}= c \cdot \dot{x}(t)
= [cM]\cdot x(t) = 0~.
\end{equation}
The existence of the conserved quantity allows to calculate the final state in
terms of the initial condition $x(0)$ as
\begin{equation}
x(\infty) := \lim_{t \rightarrow \infty} x(t) = \frac{c \cdot x(0)}{c \cdot e} e,
\label{eq:finalfrominit}
\end{equation}
where $e$ is a right eigenvector of $M$ for $\mu_{max}$.  This equation implies
that the projection of $x(0)$ on $c$ is all the system remembers at large times
about the initial condition $x(0)$. The coefficient $c_i$ quantifies the extent
to which the initial condition at node $i$ affects the final state. Therefore,
we call $c_i$ the {\em dynamical influence} (DI) of element $i$ on the dynamics under
equation~(\ref{eq:main}). 

One advantage of DI is that it is easily calculated without expensive numerical
simulations. In fact, a simple way to calculate  $c$ furthers the understanding
why this object quantifies the role of nodes in
spreading dynamics. The  power method (also called power iteration)
\cite{Mises:1929} approximates $c$ by applying higher and higher powers of $M$
to a uniform vector $w^{(0)} = (1,1,\dots,1)$. For general exponent $l \in
\mathbb{N}$, the $i$-th entry $w^{(l)}_i$ of
\begin{equation} \label{eq:walksl}
w^{(l)} = (1,\dots,1) M^l
\end{equation}
is the number of all possible walks of length $l$ departing from node $i$ or, in
other words, the number of ways an item can spread for $l$ steps when
originating at node $i$.  At the first iteration this yields
\begin{equation}
w^{(1)} = (1,\dots,1) M = (d_1,d_2,\dots,d_n)
\end{equation}
where $d_i$ is the sum over the $i$-th row of $M$.  When $M$ is the adjacency
matrix of a network, then $w^{(1)}_i=d_i$ is the (out-)degree, the number of
(outgoing) connections of node $i$. For exponent 2, the $i$-th entry $w^{(2)}_i$
is the sum of the (out-)degrees of all neighbors of $i$. This is the same as the
number of possibilities (walks) to depart from node $i$ following two links. Now
in the limit $l \rightarrow \infty$, the direction of the eigenvector $c$ is
approached by
\begin{equation}
\lim_{l\rightarrow \infty} \frac{w^{(l)}}{||w^{(l)}||} = \frac{c}{||c||}
\end{equation}
when the largest eigenvalue of $M$ is non-degenerate and larger in magnitude
than the other eigenvalues. Hence, the dynamic influence $c_i$ of element $i$ is
its ability to serve as the origin of many arbitrarily long walks on the
network.


\begin{figure}[b]
\centerline{\includegraphics[width=0.48\textwidth]{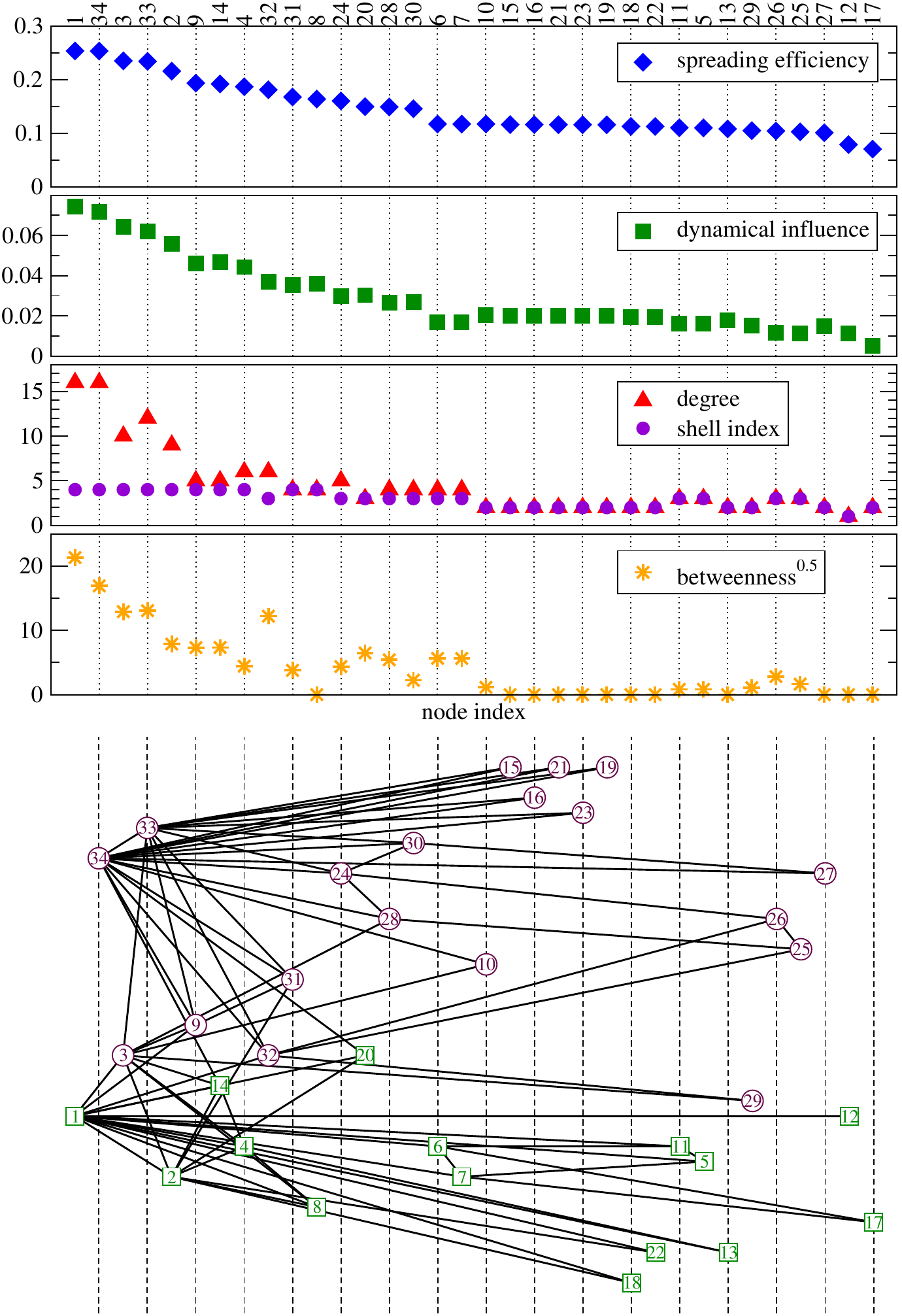}}
\caption{\label{fig:karate}
SIR spreading efficiency compared to centrality measures in a social
network. The network of Zachary's karate club \cite{Zachary:1977}
has 77 edges connecting 34 nodes, here ordered according to decreasing
spreading efficiency. A monotonic decay of a centrality measure in the
diagram indicates large predictive power for spreading efficiency.
The rank order correlation of spreading efficiency is
of 0.97 with dynamical influence, 0.86 with degree, 0.82 with shell
index, and 0.79 with betweenness centrality. Indexing of nodes is the same as
in ref.~\cite{Zachary:1977}. In the network drawing, circles and squares represent
the primary partitioning of the node set found by Girvan and Newman
\cite{Girvan:2002}. Spreading
efficiency has been estimated at $\beta =\beta_c =0.15$ performing $10^6$ independent
runs of the SIR model per seed node. The largest eigenvalue of the network
adjacency matrix is $6.65$.
}
\end{figure}

\begin{figure*}
\centerline{\includegraphics[width=\textwidth]{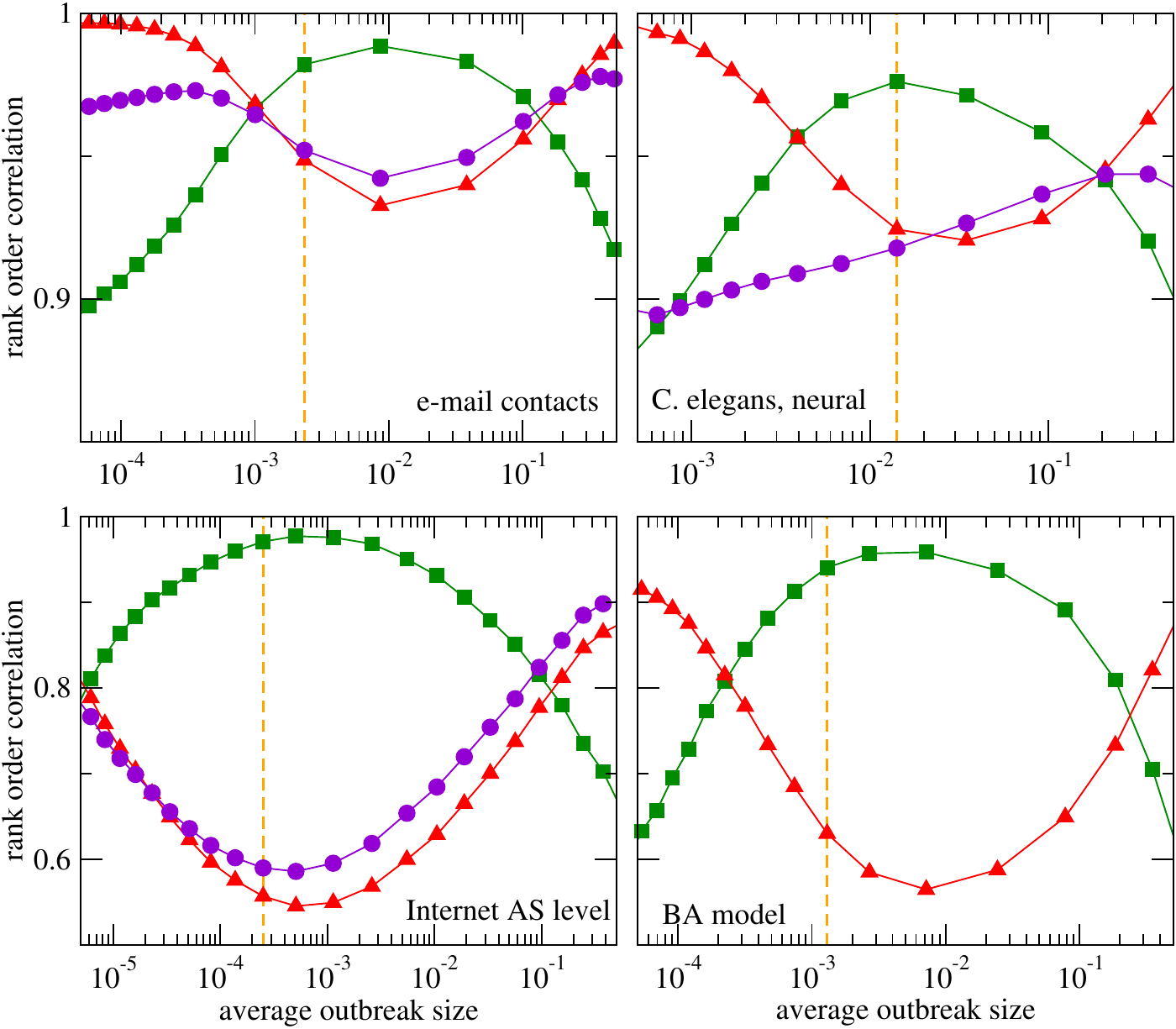}}
\caption{\label{fig:siro_combined}
Predictive power of different centrality measures for SIR spreading efficiency.
Symbols are values
of the rank order correlation coefficient of spreading efficiency with influence
(squares), degree (triangles) and shell index (circles). The choice of the
spreading parameter $\beta$ controls the average outbreak size (horizontal
axis), being the average number of nodes infected when choosing the seed node
uniformly. The vertical dashed line indicates average outbreak size at the
critical value of the spreading parameter $\beta=\beta_c$. The predictive power
of betweenness centrality is below that of degree in all cases. The following
networks have been used. E-mail interchanges between employees of a university
\cite{Guimera:2003}, $\beta_c = 0.0482$; unweighted neural circuitry of the
roundworm C. elegans \cite{Watts:1998,White:1986}, $\beta_c=0.0654$; snapshot of
the Internet at Autonomous Systems level of Nov~08,~1997, see
http://moat.nlanr.net, $\beta_c = 0.0315$; a realization of the
Barab\'{a}si-Albert (BA) model of scale-free networks  \cite{Barabasi:1999} with
1000 nodes and $m=2$ edges added per node, $\beta_c = 0.0945$. Other
realizations of the BA model yield qualitatively the same result. For the BA
model, shell index is not a predictor because its value $k_i=m$ is the same for
all nodes. For the neural network, being directed, out-degree instead of degree
is used as a predictor and for calculating the shell index.
}
\end{figure*}

\begin{figure*}
\centerline{\includegraphics[width=\textwidth]{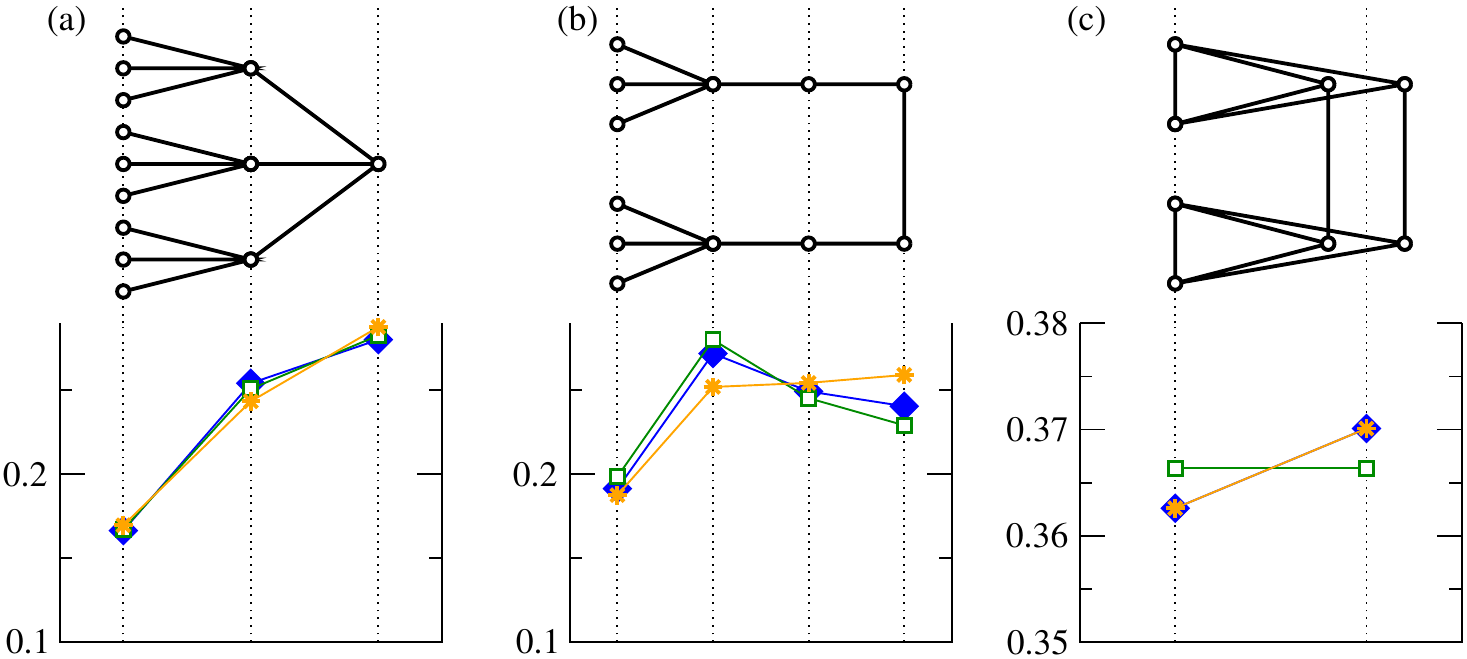}}
\caption{\label{fig:eff_illu}
Comparison between spreading efficiency (diamonds),
dynamical influence (squares) and betweenness centrality (stars)
in small networks. In panel (a), the ranking of nodes with respect to spreading
efficiency is rendered both by dynamical influence and
betweenness. Note that the most efficient spreader is not a node
with maximum degree but the node on the right connected to those
maximum degree nodes. In the case of panel (b), the strongest spreaders
are the nodes of maximum degree 3. However, the degree does not uniquely
reveal the second strongest spreaders. Dynamical influence renders the full
ranking of spreading efficiency.
In panel (c), nodes are indistinguishable both by degree and
dynamical influence. The small differences in spreading efficiency
---note the scale on the axis---on this regular graph are rendered correctly by
the betweenness centrality.
The shell index is not usable as a node discriminator here. It takes
value 1 on each node in panels (a) and (b) and the value 3 in panel (c).
Spreading efficiency is calculated at the critical value $\beta=\beta_c$
for each network, being 0.408 for (a), 0.463 for (b),and 0.333 for (c).
For easier comparison, values of dynamical influence and betweenness centrality
have been rescaled and shifted such that their mean and standard deviation
are identical to that of the spreading efficiency in each network.
}
\end{figure*}

\begin{figure*}
\centerline{\includegraphics[width=\textwidth]{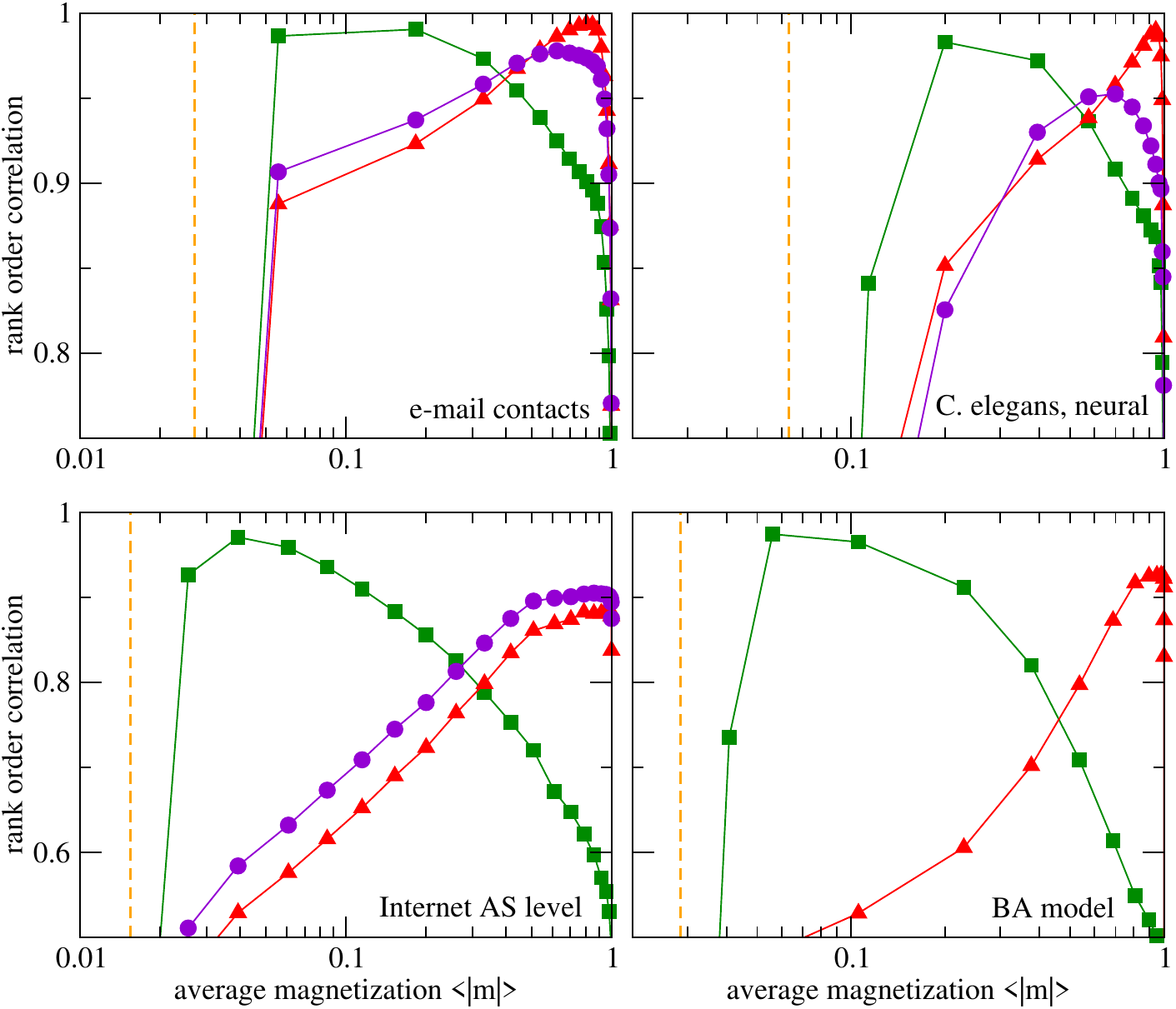}}
\caption{\label{fig:isingo}
Predictive power of different centrality measures for Ising spreading efficiency
at time lag $\tau=10$ as a function of average absolute magnetization $\langle |m| \rangle$. 
Symbols are values of the rank order
correlation coefficient of spreading efficiency with influence (squares), degree
(triangles) and shell index (circles). The vertical dashed lines indicate the 
value of $\langle |m| \rangle$ at the critical parameter value $\beta=\beta_c$.
Details on networks and the values of $\beta_c$ are given in the caption of
Figure~\protect\ref{fig:siro_combined}.}
\end{figure*}

\subsection*{Epidemic spreading} 

Let us first apply these insights to critical phenomena like spreading processes
\cite{Dorogovtsev:2008}. In the SIR model
\cite{Keeling:2007,Bansal:2007,House:2011}, each node is either susceptible,
infected or removed. An infected node $i$ transfers the epidemic along each of
its outgoing arcs independently with probability $\beta$; node $i$ itself
relaxes to the removed state at unit rate. We study small
perturbations to the stationary state with all nodes susceptible and approximate
the dynamics by the linearization
\begin{equation} \label{eq:sirlin}
\dot{x} = - x + \beta A^{\text T} x~.
\end{equation}
Here $x_j(t)$ is the probability of node $j$ to be infected at time $t$. The
first term is the relaxation from the infected to the removed state at unit
rate. The second term quantifies the transmission of the epidemic where the
network enters by the transpose of its adjacency matrix $A$.

Equation~(\ref{eq:sirlin}) can be rewritten as equation~(\ref{eq:main}) with
$M=\beta A^{\text T} - I$, and $I$ being the identity matrix. Matrix $M$ has largest
eigenvalue $\mu_{max}=0$ when the spreading probability $\beta$ is the inverse
of the largest eigenvalue of $A$, that is $\beta= \beta_c = 1 /
\alpha_{\text{max}}$. We take again $c$ as a left eigenvector of $M$ at
$\mu_{max}=0$ or, equivalently, a right eigenvector for maximum
eigenvalue $\alpha_{\text{max}}$ of $A$. Then the expected outbreak size from
an initial infection described by the probability vector $x(0)$ is proportional to
$c \cdot x(0)$.

Now we ask how well $c$ may forecast the actual SIR spreading dynamics, measured
as the {\em spreading efficiency} (details in Methods) of node $i$ that we
define as the expected fraction of nodes reached by an epidemic outbreak
initiated with node $i$ infected ({\em seed node}), all others susceptible.
Figure~\ref{fig:karate} shows that $c_i$ is a good predictor of SIR spreading
efficiency at critical parameter value $\beta = \beta_c$ in a small social
network. Dynamical influence $c_i$ outperforms the predictions made by degree,
shell index and betweenness centrality. Predictive power is quantified by the
rank order correlation (see Methods).

Figure~\ref{fig:siro_combined} shows the predictive power of dynamical influence
for spreading efficiency as a function of the infection probability in larger
real-world networks and the Barabasi-Albert model. The results are as
anticipated by the theory. Dynamical influence is a good predictor of spreading
efficiency in the critical regime where $\beta / \beta_c \approx 1$. Predictions
by dynamical influence outperform those by other quantities that are supposed to
provide information about expected outbreak size in a broad interval of
infection probabilities. This still holds for values of $\beta$ that lead to
average outbreak sizes of up to 10\% of all nodes in the network, as indicated
by the vertical dashed lines in Figure~\ref{fig:siro_combined}.

The approximation $w^{(l)}$ for {\em finite} length $l$ in Eq.~(\ref{eq:walksl})
is useful as a predictor of spreading efficiency as well. Even when the
interaction network is not completely available, local information counting the
number of walks of length $l=2$ or $l=3$ emanating from a node is enough to
estimate dynamical influence. Figure~S1 in the Supplementary Information shows that
the count of these short walks yields a prediction of spreading efficiency in
the critical regime that is as good as dynamical influence itself. The
predictive power of these walk counts, too, reaches a maximum in the critical
regime.

For infection probabilities $\beta$ far above or below the critical value
$\beta_c$, however, the degree $d_i$ of a node $i$ is a better predictor of
spreading efficiency. In the subcritical regime, spreading is sparse and
typically confined to the neighborhood of the seed node $i$,  while in the
supercritical regime, the epidemics rarely fails to spread to the whole system.
In the critical regime in-between these extremes, infectious seeds are
perturbations that trigger relaxation dynamics at all scales. This is reflected
in a dynamics dominated by a marginal linear mode and a variety of possible
final states. Dynamical predictions at criticality require then a global view of
the network structure (and the final state is determined by the conservation law
associated with the leading eigenvector $c$-removed). The scale-free
distribution of epidemic outbreaks in real populations
\cite{Rhodes:1996,Pinto:2011} is a sign of criticality and suggests that this
regime is most relevant in practice.

In order to check the robustness of the results we consider two modifications of
the epidemic spreading dynamics. First, we study the SIR model with a stochastic
rather than deterministic transition from the infected to the removed state.
Specifically, the transition occurs with a probability $\mu$ independently for
each infected node at each time step. Thus the time spent in the infected state
(recovery time) has a geometric distribution with mean $\mu^{-1}$. This
modification does not qualtitatively change the results of Figure~\ref{fig:siro_combined}
up to rescaling of $\beta$ with $\mu$. In fact, the
curves of predictive power for different values of $\mu$ collapse when plotted
as a function of the average outbreak size, see Figure~S2. Second, prediction by
dynamical influence may also be applied to the SIS model (see Materials and
methods) yielding very similar results (Supplementary Information, Figure~S3). 
The contact process \cite{Harris:1974} can also be considered, with $A$
replaced by the stochastic adjacency matrix, the adjacency matrix after
normalization such that each row sums up to 1.

To facilitate intuitive understanding of the predictive role of centrality
measures in spreading dynamics, let us consider small networks. In each
of the three cases in Fig.~\ref{fig:eff_illu}, a different subset
of the measures yields the correct ranking by spreading efficiency at the
critical point. The most efficient spreader is not necessarily the node with the
largest degree. Being adjacent to several nodes with large degree may lead to
large spreading efficiency despite a smaller degree, cf.\ panel (a). This second
order effect is reflected by dynamical influence. When all degrees are equal as
in panel (c), also dynamical influence and shell index are homogeneous. In this
case, betweenness centrality captures the subtle effect of nodes having
different positions in the network. We speculate that centrality measures based
on unconstrained walks and shortest paths can do best in predicting spreading
efficiency at the critical point. Then, a suitable combination of dynamical
influence with betweenness depending on network topology might be close to the
optimal predictor.

Nodes with large shell index are contained in highly connected neighborhoods
that facilitate spreading. In many cases, the shell index may serve as a satisfactory
predictor of spreading efficiency \cite{Kitsak:2010}. Here, however, we find situations where
its use for prediction is limited due to the degeneracy of the values shell index assumes. 
It has a constant value $m$ across nodes
of each network that can be built up by iterative attachment of a node with
exactly $m$ edges. This includes all trees ($m=1$) and networks from growth models
such as the one by Barabasi and Albert \cite{Barabasi:1999}. A significant lack of resolution is
also observed in real-world networks. Shell index
assumes only few ($\approx 10$) discrete values, cf.\ Figure~S4 in Supplementary Information.
On the Internet graph, the same maximum shell index value is observed for the strongest
spreader as well as nodes with spreading efficiency a factor of five below. Thus, even though
the overall correlation between spreading efficiency and shell index is positive,
lack of resolution limits the predictive power. Such limitations have been identified also in an
empirical study of epidemic spreading in a social group \cite{Christakis:2010}, in a
detailed comparison between SIR and SIS models \cite{Castellano:2011} and in dynamics
of rumour propagation \cite{Borge-Holthoefer:2011}.

We remark that the most efficient spreaders are not necessarily the same as
those targeted by efficient vaccination strategies in order to contain
epidemics. At the network level, the aim of vaccination is to increase
the epidemic threshold $\beta$ in order to render the spreading
dynamics subcritical. The set of nodes by whose removal this shift of
threshold is achieved \cite{Restrepo:2006} is different in general from
the set of nodes with the largest dynamical influence. The Supplementary Information
provides further results (Figure~S5) and a brief discussion of vaccination.


\subsection*{Ising model}

The Ising model is a paradigmatic binary state model of critical phenomena. The
Ising model \cite{Ising:1925,Brush:1967} on a network \cite{Dorogovtsev:2002b}
describes the dynamics of $N$ coupled spins $s_i \in \{-1;+1\}$ placed on the
nodes. The zero temperature (T=0) version of the Ising model implements a
majority rule for state updating. This is the same dynamics considered in
threshold models of collective behavior for a $50\%$ value of the threshold
\cite{Granovetter:1978,Gonzalez-Avella:2011}, and its dynamics is also related
to Schelling's model of urban segregation \cite{Schelling:1971,Vinkovic:2006};
the finite temperature version has been considered in the context of strategic
interactions \cite{Blume:1993}. Finite temperature effects (noise), as
considered here, are essential to escape from frozen configurations and to
establish the robustness of transitions found in Ising-like models
\cite{Klemm:2003}. Also in the theory of neural computation,
Ising-like systems play an essential role \cite{Hertz:1991}.

In the context of the Ising model, we define spreading efficiency of
node $i$ as the correlation between two measurements: the state of
node $i$ at time $t$ and the magnetization (see Methods) of the whole
system at a later time, formally
\begin{equation}
f_i(\beta) = \langle s_i(t) m(t+\tau N) \rangle~.
\end{equation}
The parameter $\tau$ measures the time lag between the two measurements.
Figure~\ref{fig:isingo} shows to which extent the ranking of nodes by Ising
spreading efficiency is correlated with the ranking by various centrality
measures. At the transition between order and disorder, Ising spreading
efficiency has larger correlation with dynamical influence than with the
other centrality measures.


\subsection*{Diffusive processes: the voter model}

Coming back to the general framework equation~(\ref{eq:main}),
there is a class of dynamical processes in networks in which the property of $M$
having a zero maximum eigenvalue appears naturally without the need of adjusting
any parameter. This is the case of diffusive processes defined by
equation~(\ref{eq:main}) with $M=-L$ and the Laplacian matrix entries
\begin{equation} \label{eq:diffusion}
L_{ij} = - K_{ij} + \delta_{ij} \sum_{k=1}^N K_{ik}~.
\end{equation}
The zero eigenvalue of $L$ is non-degenerate under mild assumptions
\cite{Agaev:2005}. For these processes our general analysis of
equation~(\ref{eq:main}) becomes exact. A prominent example of  diffusive
dynamics is the voter model \cite{Holley:1975} in which node $i$ is in a spin
state $s_i \in \{-1,+1\}$. For this model, $x_i$ stands for the ensemble average
of spin $i$,  $x_i = \langle s_i \rangle$, and $K_{ij}$ gives the rate at which
node $i$ copies the state of node $j$. Different definitions of the voter model
dynamics provide clear examples of how the concept of dynamical influence takes
into account the interplay between topology and dynamics: For {\em link update}
dynamics in an undirected network, an ordered pair of nodes $(i,j)$ is chosen in
each step and node $i$ copies the state of node $j$. The rate matrix $K$ becomes
proportional to the transpose of the adjacency matrix $A$. As a consequence
$c_i=1/N$, the average magnetization $\sum_{i=1}^N c_i x_i$ is conserved, and
all nodes have the same dynamical influence independently of the topological
features of the network. In the more standard {\em node update} voter dynamics,
at each step one node $i$  (having degree $d_i$) is selected at random and
copies the state of one of its neighbors $j$, also selected at random. In this
case $K_{ij} \propto A_{ji}/d_i$, so that $K_{ij}$ is no longer a symmetric
matrix, the conserved quantity is a weighted magnetization \cite{Suchecki:2005a}
and the dynamical influence of node $i$ is proportional to its degree $d_i$.

For diffusive processes, the system is driven towards a homogeneous final state
with $x^\ast: = x_i(\infty)= x_j(\infty)$ for all $i$ and $j$. Although $x^\ast$
takes continuous values, each realization of the voter dynamics in a finite
system eventually reaches a homogeneous absorbing state with either all nodes in
the state $+1$ or all in the state $-1$. The influence $c_i$ of a node weights
the initial state of node $i$ in the exit probability $P_+$, that is, the
probability to reach the absorbing configuration $+1$: $P_+ =(\phi_ c +1)/2
=(x^\ast +1)/2$. When all nodes are equivalent (e.g., link update) $x^\ast$  is
just the average of the initial values of the nodes, but otherwise (e.g. node
update) $x^\ast$ is given by a weighted average of the initial condition.
The value $c_i$ has an alternative interpretation as a stationary density of a
random walk \cite{Masuda:2010}.

\begin{figure}
\centerline{\includegraphics[width=0.48\textwidth]{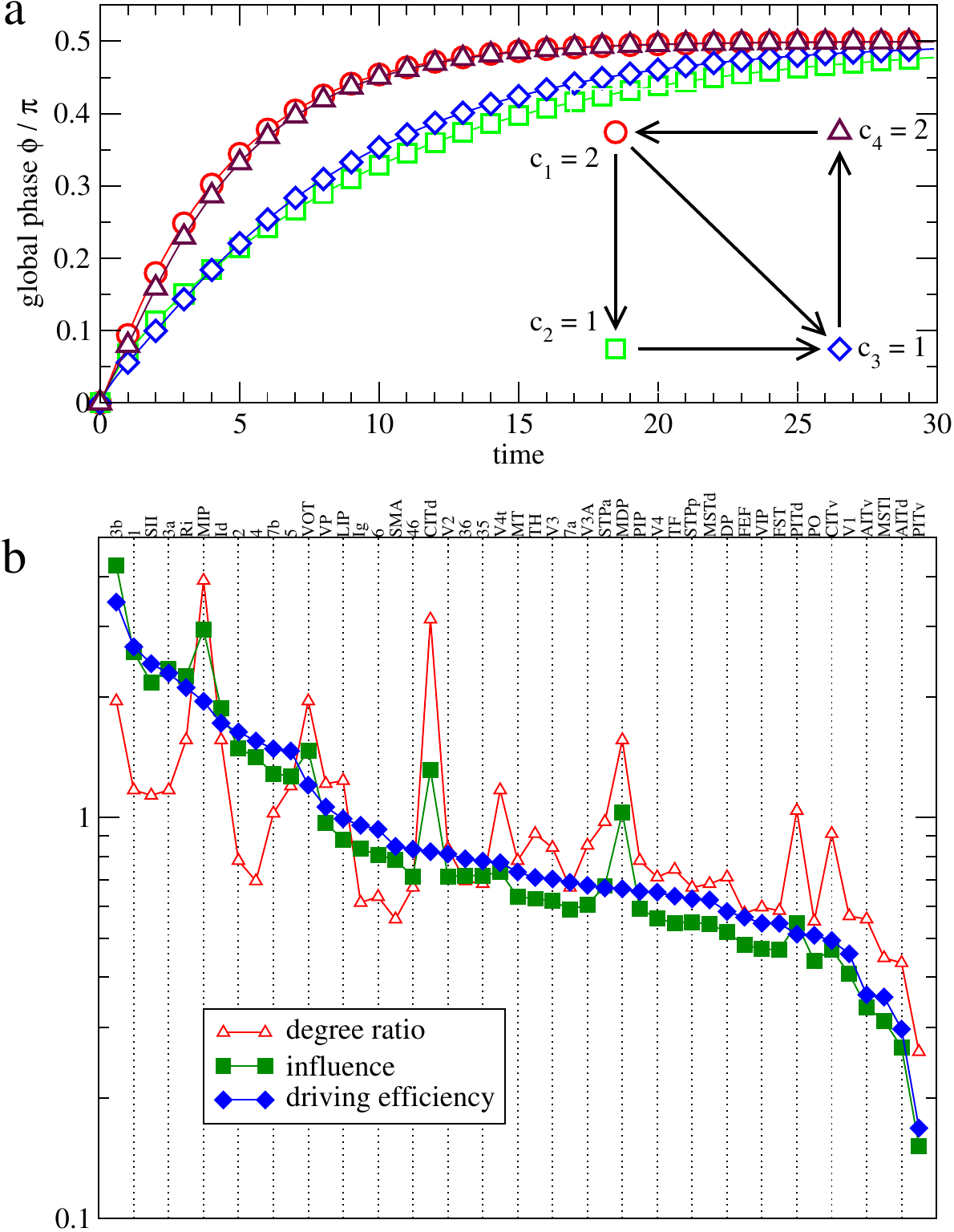}}
\caption{\label{fig:panel4} Dynamical influence and driving in a system
of phase-coupled oscillators. {\bf a} Adaptation of the {\em global} phase of
the four-node system in the inset when driving is applied to one of the
oscillator nodes (an additional oscillator coupled to the red circle
node with fixed phase $\pi/2$. Adaptation is quick when driving at
nodes 1 ($\circ$) and 4  ($\bigtriangleup$), having high influence, and slow
when driving at one of the other two nodes, having low influence. {\bf b,}
Driving efficiency (filled diamonds) and dynamical influence (shaded squares)
for a system of phase oscillators coupled as the network of regions in the macaque
cortex \protect\cite{Honey:2007}. Driving efficiency of node $i$ is measured as
the time required to resynchronize with an additional input signal applied to a
given node $i$.  We say that the system has resynchronized when the global
phase reaches $\psi(t) \ge (1-\epsilon) x_a$ with a tolerance $\epsilon =
10^{-2}$, where the global phase $\psi(t)$ is computed as the argument of the global
order parameter $z(t)=r \exp[i\psi(t)]= \sum_{i=1}^N \exp[\imath s_i(t)]$.
For comparison, the degree ratio $k_{\rm out} / k_{\rm in}$ of
each node is also shown (open triangles). Since this is not an uncorrelated
network, node influence deviates significantly from degree ratio. Each plotted
quantity is rescaled by a factor to obtain an average value of 1. The
empirical network serves as a testbed for prediction of driving efficiency.
We do not aim to mimic real dynamics of the cortex.
}
\end{figure}

\subsection*{Efficient driving of complex systems} 
The meaning of dynamical 
influence also manifests itself in the practical task
of driving a system efficiently. In the context of the voter model, this task
might be phrased in terms of the zealot problem \cite{Mobilia:2003}.
One considers a
special directed network in which a given node (the zealot) does not copy the
state of any of its neighbors. The question is the efficiency of the zealot in
driving all other nodes to the zealot state.  To show the broad applicability of
the dynamical influence concept, we address this question of driving efficiency
considering the problem of phase-coupled oscillators described by the Kuramoto
model \cite{Acebron:2005}. Assuming all oscillators have the same intrinsic
frequency $\omega$ (without losing generality, we choose
$\omega=0$), the phase variable $x_i$ of oscillator $i$ advances as
\begin{equation} \label{eq:nonlinosci}
\dot{x_i} = \omega + \sum_{j=1}^N K_{ij} \sin(x_j - x_i)~.
\end{equation}
with a matrix $K$ of non-negative coupling strengths. Around the synchronized
state, phase differences are small. By approximating each $\sin$-term with its
argument, a linear homogeneous system as in equation~(\ref{eq:main}) is
recovered.

We study a scenario with initially all oscillators $i$ in phase $x_i(0)=0$. An
additional node $a$ with constant phase $x_a = \pi /2$ is added to the system
and linked through an additional edge to a chosen node $i$. We measure the time
$T_i$ the system takes to reach the new homogeneous state with $s_i = \pi/2$ for
all nodes $i$. The dynamical evolution of these systems is illustrated by
studying the motif in the inset of Fig.~\ref{fig:panel4}a. The global phase
$\psi(t)$ converges faster to the external forcing when the driving is applied
to the nodes with higher influence, and the convergence of the different nodes
depends on their relative network position in relation to the driver. In
Fig.~\ref{fig:panel4}(b), we show the results on a directed network of phase
oscillators connected as the network of regions in the macaque cortex
\cite{Honey:2007}. Dynamical influence has extremely high predictive power. The
rank order correlation of driving efficiency with dynamical influence is $0.97$,
while $0.66$ with degree ratio, $-0.14$ with shell index and $-0.09$ with betweenness.
Similar results are obtained on randomly grown directed networks
and for coupled chaotic oscillators, see Figures~S6 and S7 in Supplementary Information.
These findings clearly show that dynamical influence is an excellent proxy to identify
better targets for controlling global behavior, even in strongly non-linear dynamical
systems.


\section*{\large Discussion}

Dynamical influence is a centrality measure applicable to a wide range of
dynamical processes on complex networks that takes into account the interplay
between topology and dynamics. While the motivation and rigorous analysis of
dynamical influence employ the context of linear systems, its practical use for
understanding and controlling networked dynamics extends to several inherently
non-linear systems.

We have demonstrated that dynamical influence is applicable to stochastic
equilibrium (Ising model) and non-equilibrium systems (epidemic and voter
models) as well as deterministic state-continuous systems such as the Kuramoto
model and the chaotic Roessler attractor. For critical epidemic spreading and
the Ising model, dynamical influence is a good predictor of spreading
capabilities. In the context of chaotic Boolean dynamics \cite{Kauffman:1993}, a
similar spectral centrality is highly correlated with a node's impact on the
attractor reached \cite{Ghanbarnejad:2011}. For diffusion, dynamical influence
quantifies the impact of the dynamical states of single nodes on the asymptotic
homogeneous state. Beyond that, it proves to be a high-quality proxy for driving
efficiency, uncovering which are the best target nodes in real networks to be
forced in order to drive the system towards specific states. In a broader
context, the identification of these targets has fundamental implications and
practical applications on strategies with an interest in controlling collective
behavior, from social influence to biomedical responses.


\section*{\large Methods}

\subsection*{Epidemic models}

We simulate the {\em SIR model} of epidemic spreading in the time-discrete
version. Transitions between the three states (S,I,R) are as follows.  If node
$i$ is in the S (susceptible) state and has $\nu$ infected (I) neighbors at
time $t$, then node $i$ remains susceptible with probability $(1-\beta)^\nu$,
otherwise $i$ is infected at time $t+1$. If node $i$ is in the infected state at
time $i$ then $i$ is in the R (removed) state at time $t+1$. In the {\em SIS
model}, at difference with SIR, a node infected at time $t$ is susceptible again
at time $t+1$. The probability of being removed in the {\em SIR model} does not
enter in the linearized equation~(\ref{eq:sirlin}) because it appears only in a
second order term in the equation for $x$. Therefore equation~(\ref{eq:sirlin})
gives the same linear description for the {\em SIR} and {\em SIS} models.

The system is in an absorbing configuration if none of the nodes is infected.
For both models, outbreak size is the number of nodes having been infected at
least once before reaching an absorbing configuration. The spreading efficiency
of node $i$ is the average outbreak size when initiating the dynamics with node
$i$ infected and all others susceptible.

\subsection*{Ising model}

The spin values $s_i \in \{-1,+1\}$ are updated asynchronously
as follows. At each time step $t$, a node $i\in\{1,\dots,N\}$ is drawn
uniformly. The field
\begin{equation}
h_i(t) = \sum_{j=1}^N K_{ij} s_i(t)
\end{equation}
is calculated. The state of node $i$ is flipped with probability
\begin{equation}
\min \{ \exp[- \beta s_i(t)h_i(t)] ,1 \}~.
\end{equation}
Flipping the state of node $i$ means $s_i(t+1)=-s_i(t)$. Otherwise
the state of node $i$ remains unchanged. All other nodes $j \neq i$ retain
their state, $s_j(t+1)=s_j(t)$.

The parameter $\beta$ (inverse temperature) controls the order in the
system. For large $\beta$ (small temperature), spins tend to align and there is
long-range order seen as large clusters of nodes sharing the same spin value.
For small $\beta$ (high temperature), long-range order is absent.
The magnetization
\begin{equation}
m(t)= N^{-1} \sum_{i=1}^N s_i(t)~.
\end{equation}
is used to quantify the order of the system.
Disordered systems have $\langle |m| \rangle \approx 0$, while a finite
positive value is obtained in ordered systems.

\subsection*{Rank order correlation}

For a vector $x \in \mathbb{R}^n$, the rank of component $i$ is given by
\begin{equation}
r_i(x) = 1 + | \{ j \neq i | x_j > x_i \}| +
\frac{1}{2}|\{j \neq i | x_j = x_i \}|
\end{equation}
The rank order correlation coefficient $\rho(x,y)$ between two such vectors
$x$ and $y$ is the Pearson correlation coefficient between the rank vectors
$r(x)$ and $r(y)$. Thus $\rho(x,y)$ takes values in $[-1,1]$ with
$\rho(x,y) = +1$ $(-1)$ if and only if $x$ and $y$ are in a strictly
increasing (decreasing) relation.

\subsection*{Degree and degree ratio}

The degree $d_i$ of node $i$ is the number of nodes $i$ is connected to.
In directed networks, in- and out-degree $d_i^\text{in}$ and $d_i^\text{out}$
are distinguished. For the matrix averaging over all adjacency matrices of
networks with fixed node degrees, $c_i=d_i$ is a left eigenvector for the
largest eigenvalue. Likewise, the degree ratios
$d_i^\text{out} / d_i^\text{in}$ form a left eigenvector of the Laplacian
matrix averaging over all networks with given node degrees \cite{Serrano:2009,
Masuda:2009a}.

\bibliography{invoter_2}

\section*{Acknowledgments}
K.K.\ acknowledges financial support from VolkswagenStiftung and from European
Commission NEST Pathfinder initiative on Complexity through project SYNLET
(Contract 043312). M.A.S.\ acknowledges financial support by the Ram\'on y Cajal
program of the Spanish Ministry of Science, MICINN Project No.
BFU2010-21847-C02-02, and Generalitat de Catalunya grant No.\ 2009SGR1055.
V.M.E.\ and M.S.M.\ acknowledge financial support from MICINN (Spain) and FEDER (EU)
through projects FISICOS (FIS2007-60327) and MODASS (FIS2011-24785). 

\setcounter{figure}{0}

\clearpage
\section*{SUPPLEMENTARY MATERIAL}

\makeatletter 
\renewcommand{\thefigure}{S\@arabic\c@figure}
\renewcommand{\theequation}{s\@arabic\c@equation}
\makeatother

\begin{figure*}
\begin{center}
\centerline{\includegraphics[width=1.00\textwidth]{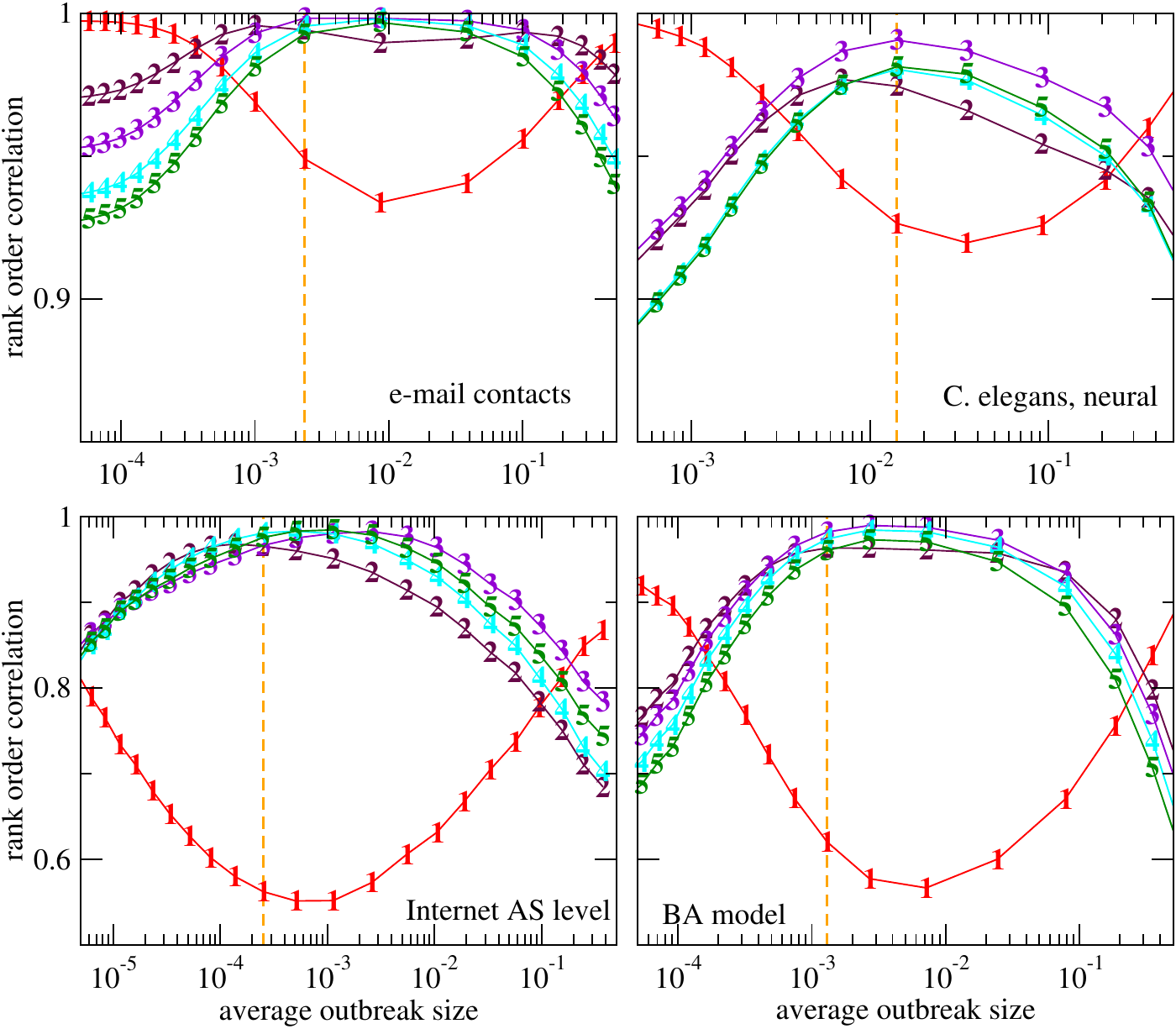}}
\caption{\label{fig:walks} 
Prediction of SIR spreading efficiency of a node $i$ by the number of walks
$w_i^{(l)}$ of a given length $l$ and starting at $i$. Symbols indicate
the value of length $l$. For $l=1$, the number of walks is the
(out-)degree. Further details are identical to those of Fig.~2 in the main article.
}
\end{center}
\end{figure*}

\begin{figure*}
\begin{center}
\centerline{\includegraphics[width=1.00\textwidth]{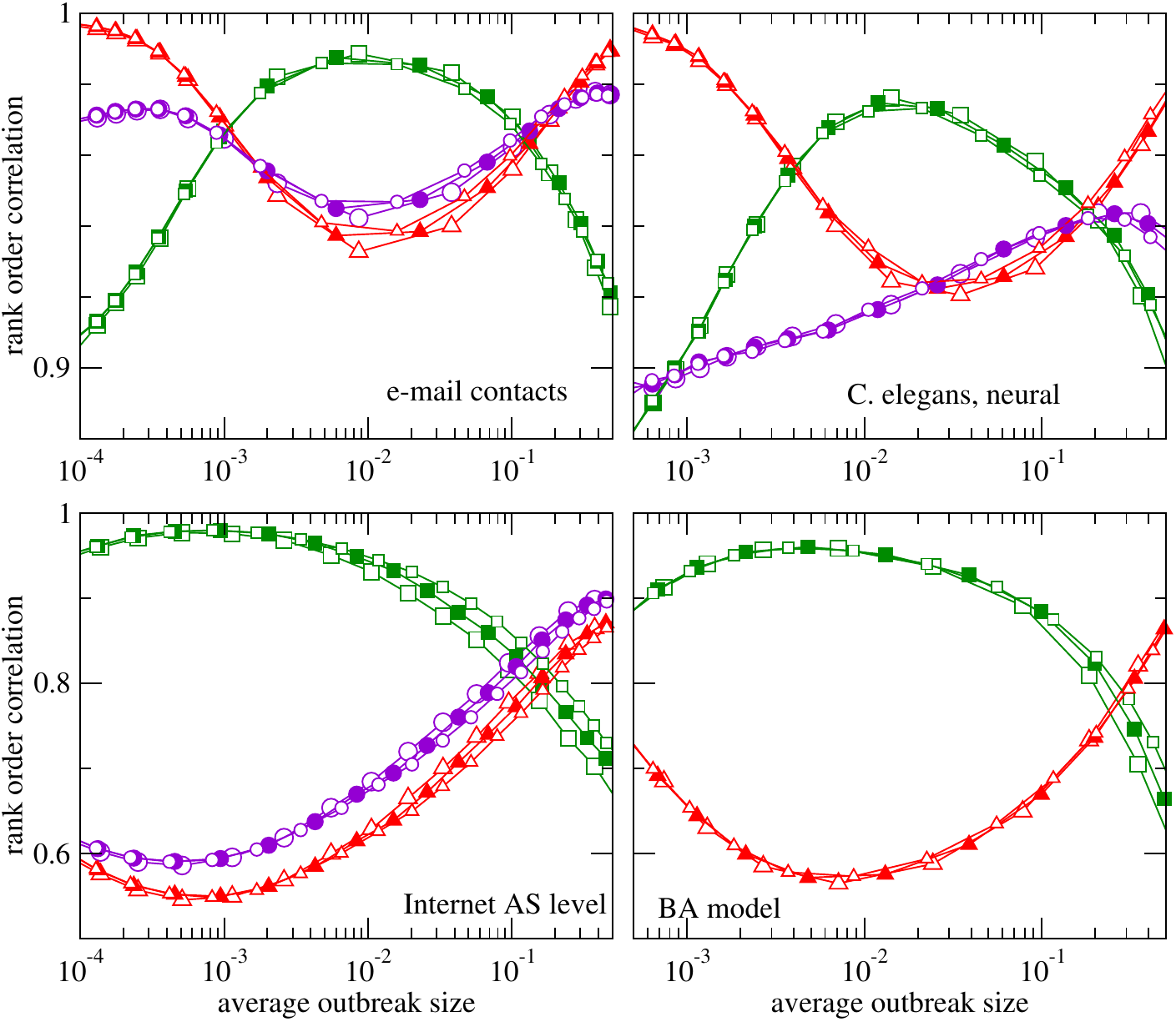}}
\caption{\label{fig:supp_figure_2} Predictive power of different centrality
measures for SIR spreading efficiency under varying recovery probability $\mu$,
distinguishing by small open symbols ($\mu=0.1$), closed symbols ($\mu=0.5$) and
large open symbols ($\mu=1.0$). Predictors are dynamical influence (squares),
degree (triangles) and shell index (circles). The x-axis is the average outbreak
size when seed nodes are chosen with equal probability. Further details are as
in Figure~2 in the main article.
}
\end{center}
\end{figure*}

\begin{figure*}
\begin{center}
\centerline{\includegraphics[width=1.00\textwidth]{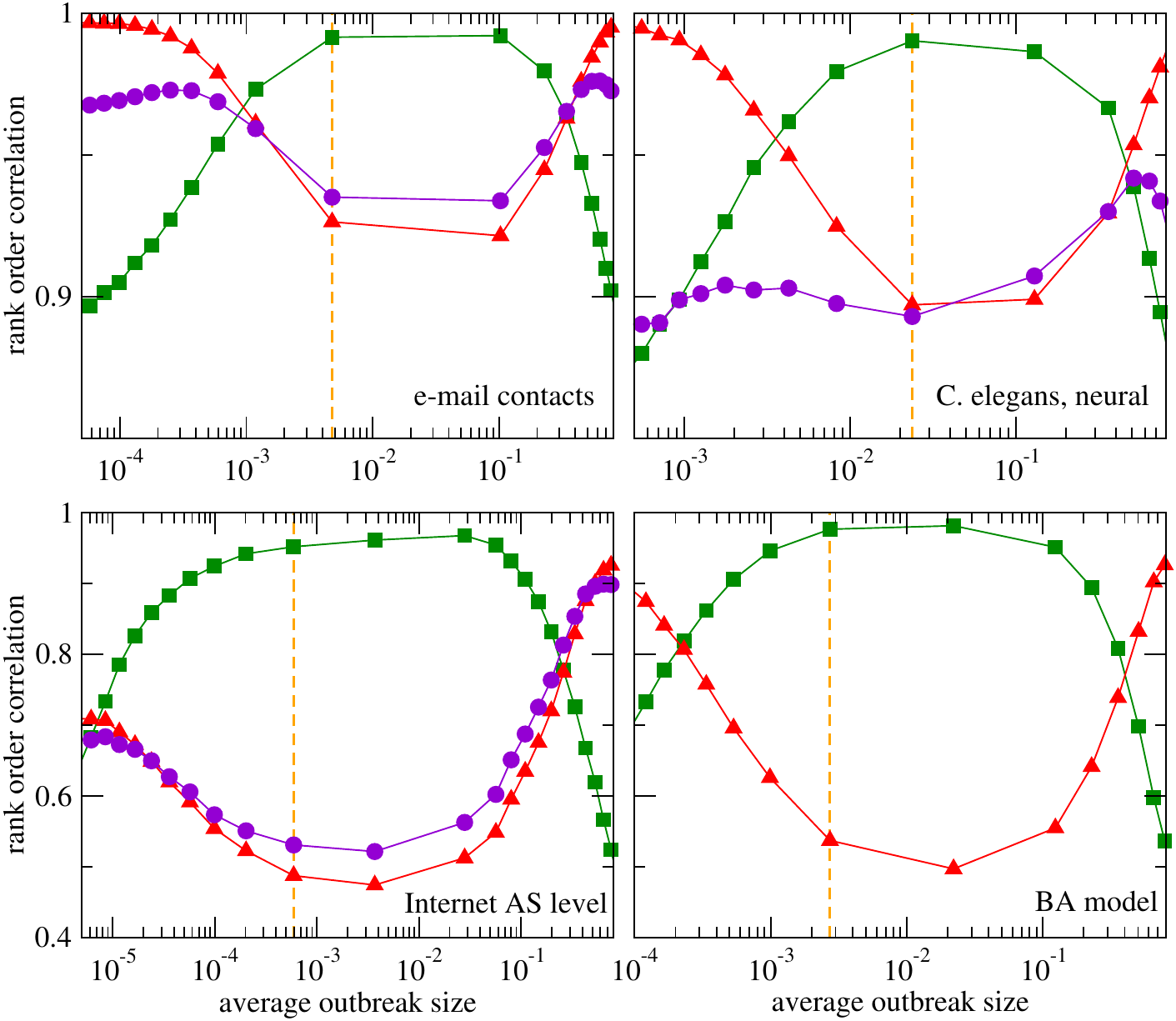}}
\caption{\label{fig:siso_combined} 
Predictive power of different centrality measures for SIS spreading efficiency.
Symbols are values of the rank order correlation coefficient of spreading
efficiency with influence (squares), degree (triangles) and shell index
(circles). Further details are as in Figure~2 in the main article.
}
\end{center}
\end{figure*}

\begin{figure*}
\centerline{\includegraphics[width=\textwidth]{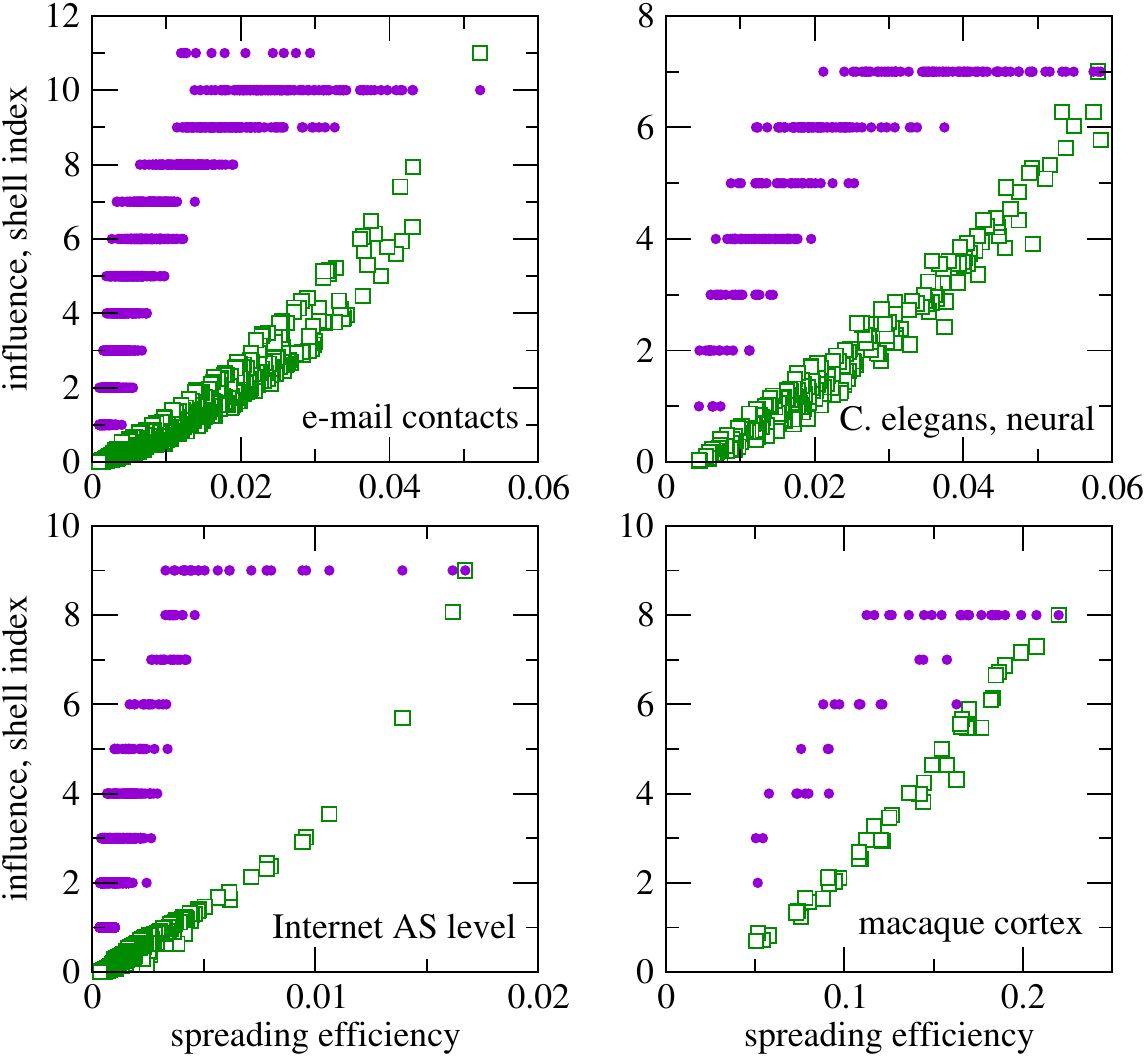}}
\caption{\label{fig:sir_scatter}
Scatter plots of the shell index (filled circles) and influence (open squares) versus spreading efficiency
in the critical regime. Each data point corresponds to a single node $i$ in the given network with the
abscissa giving the average outbreak size when seeding the SIR model at $i$. Parameter values are
$\beta=0.06$ (e-mail contacts), $\beta=0.07$ (C.\ elegans), $\beta=0.04$ (Internet), $\beta=0.09$
(macaque cortex). In each of the four cases, the normalization of the influence vector $c$ is chosen
such that its maximum equals the maximum shell index $k_{\rm max}$.
}
\end{figure*}

\begin{figure*}
\begin{center}
\centerline{\includegraphics[width=\textwidth]{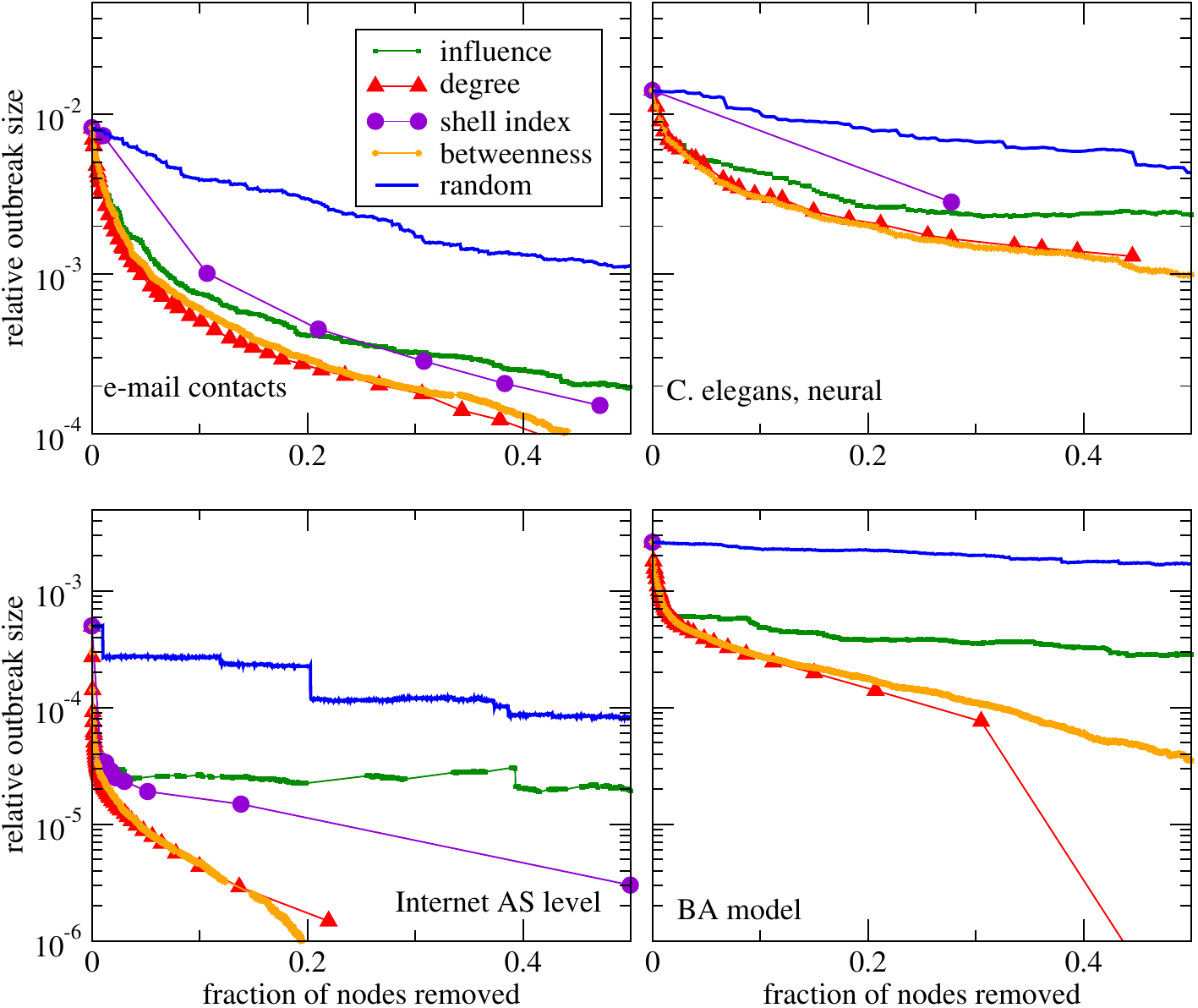}}
\caption{\label{fig:vaccination} 
Effect of node removal (vaccination) in the order of decreasing centrality
on the average size of
epidemic outbreaks in the SIR model at criticality. Partially vaccinated
networks are obtained by removing the nodes with centrality value above
a given threshold. Relative outbreak size is the average fraction of the
remaining nodes reached by an infection. Each data point is an average over
$10^6$ realizations of the SIR process with seed nodes drawn uniformly
and independently across realizations.
In each network, the spreading parameter is chosen as
$\beta=1/\lambda_\text{max}$, the inverse of the largest eigenvalue
of the adjacency matrix. When $\beta$ is two times or five times this value,
qualitatively the same result is obtained. Note that centrality measures
with discrete values (degree and shell index) produce fewer data points,
because they assign the same value to larger sets of nodes.
}
\end{center}
\end{figure*}

\subsection*{Epidemic spreading}

We test the idea of approximating dynamical influence by applying a finite ($l$-th)
power of the adjacency matrix to a uniform vector, cf.\ section {\em Defining
dynamical influence} of the main article. The resulting quantity is $w_i^{(l)}$,
the number of walks of length $l$ departing from a node. Figure~\ref{fig:walks}
shows that the number of walks of length $l\ge 2$ is a good predictor of
spreading efficiency in the critical regime, even for moderate $l$.

Further results on epidemic spreading to support the robustness of
dynamic influence as a predictor for spreading efficiency.
Figure~\ref{fig:supp_figure_2} compares results for the SIR model with different
recovery probability $\mu$. At each time step, an infected (I) node changes
state to recovered (R) with probability $\mu$ and stays infected with 
probability $1-\mu$. The case of deterministic recovery in a single step
($\mu=1$) is the one treated in Figure~2 in the main article. In
Figure~\ref{fig:siso_combined}, results for the SIS model are presented
analogous to those for the SIR model, cf.\ Figure~2 in the main article. 

Figure~\ref{fig:sir_scatter} provides the detailed relation between
centrality measures (shell index, influence) and spreading efficiency at
criticality for the standard SIR model ($\mu=1$). In the network of
e-mail contacts, the maximum shell index $k_{\rm max} =11$ is not
assumed by any of the top spreaders. In the neural network of
C.\ elegans, $k_{\rm max}=7$ is assumed by the 76 top spreaders
out of the 274 nodes. Across these 76 nodes, spreading efficiency
varies by a factor of 2.75. Likewise, the 26 top spreaders out of the
3015 autonomous systems of the Internet graph have $k_{\rm max} =9$,
with spreading efficiency varying by a factor of more than five across these
26 nodes. Finally, the macaque cortex network has 47 nodes in total;
spreading efficiency varies by a factor of 1.96
across the 23 top spreaders belonging to the innermost shell of
$k_{\rm max}=8$.

Furthermore, we study a scenario of epidemic spreading with a fraction of the
nodes vaccinated and thus immune. For these purposes, the vaccinated
nodes and their connections are removed from the network before running the SIR
dynamics. In order to test the suitability of a centrality measure for defining
a vaccination strategy, the removed nodes are those ranking highest under the
centrality measure. Figure~\ref{fig:vaccination} shows that outbreak size
is reduced most when vaccinating the nodes with the largest degree or the
largest betweenness centrality. Shell index and dynamical influence provide
less guidance when selecting candidates for vaccination. All four centrality
measures perform better than a purely random assignment of centrality values.

Why is the node set to be chosen for efficient vaccination not the same as the
group of efficient spreaders at criticality? Removing central nodes is expected
to drastically reduce the largest eigenvalue of the network and thereby increase
the epidemic threshold. Very few node removals may render the dynamics
subcritical on the network of remaining nodes. Once in the subcritical regime,
the degree is the centrality measure that best predicts spreading efficiency.
Thus for vaccination, nodes are to be selected by degree rather than dynamical
influence.

\begin{figure*}
\centerline{\includegraphics[width=\textwidth]{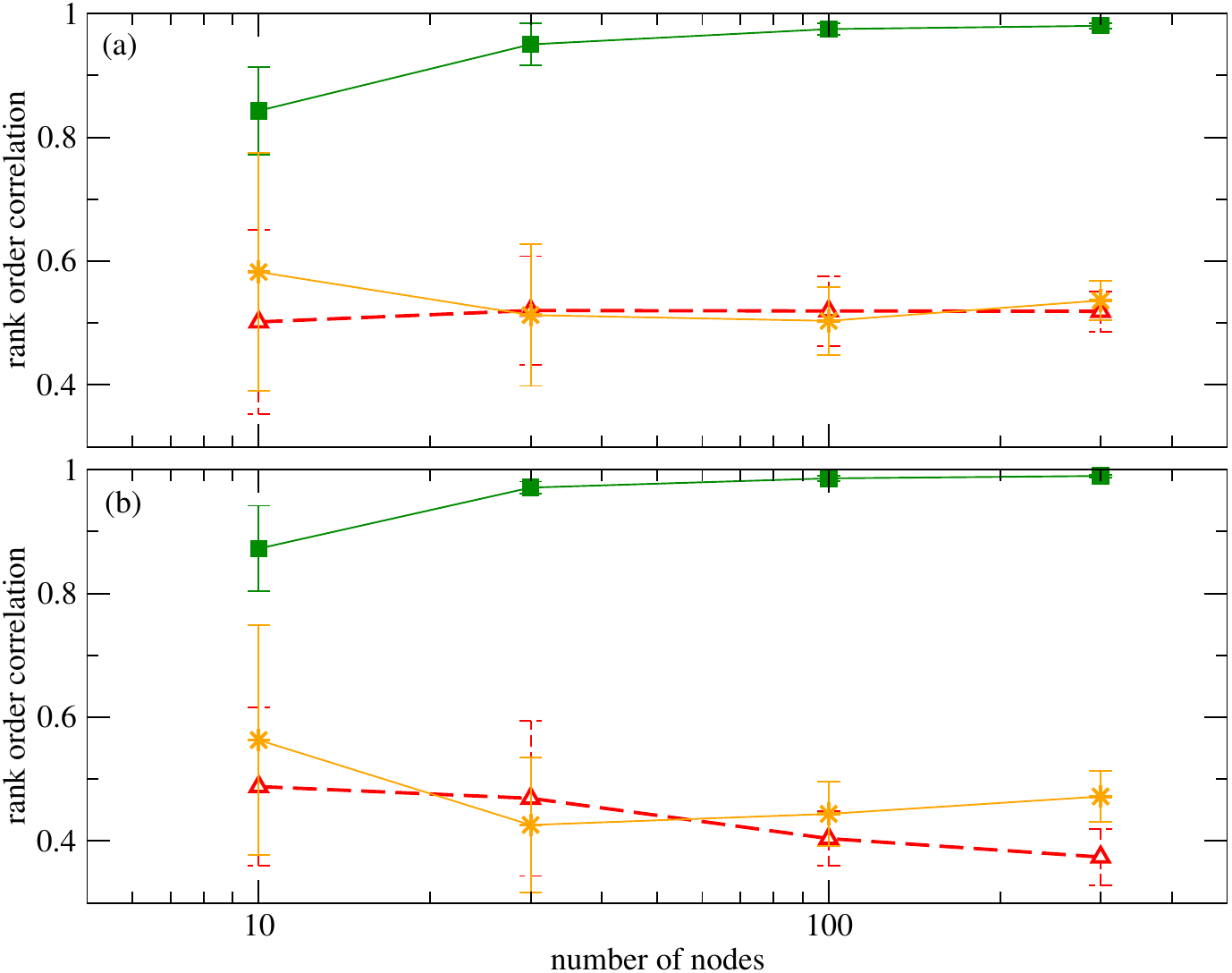}}
\caption{\label{fig:rhorscc}
Prediction of driving efficiency
for Kuramoto oscillators on randomly grown directed networks
of varying size. Each plotted point is the average rank order correlation
between driving efficiency and Laplacian influence (squares),
degree ratio (triangles), betweenness centrality (stars). Averages are
taken over 10 independently grown networks. Error bars indicate the standard
deviation. For panel (a), networks are initiated as two bidirectionally
coupled nodes. Then at each step $i$ of the growth process, a new 
node $i$ attaches to the network
with a link to a node $j$ and another link from a node $k$, where $j$ and
$k$ are chosen uniformly at random from the set of $i-1$ existing nodes.
For panel (b), the process of (a) is altered as follows. The choice of
$k$ is no longer uniform but node $k$ is chosen with probability proportional
to the out-degree of node $k$. This results in a scale-free out-degree
distribution while keeping the in-degree distribution geometric.
For the networks of panel (a), both in- and out-degree are distributed
geometrically.
}
\end{figure*}

\begin{figure*}
\centerline{\includegraphics[width=\textwidth]{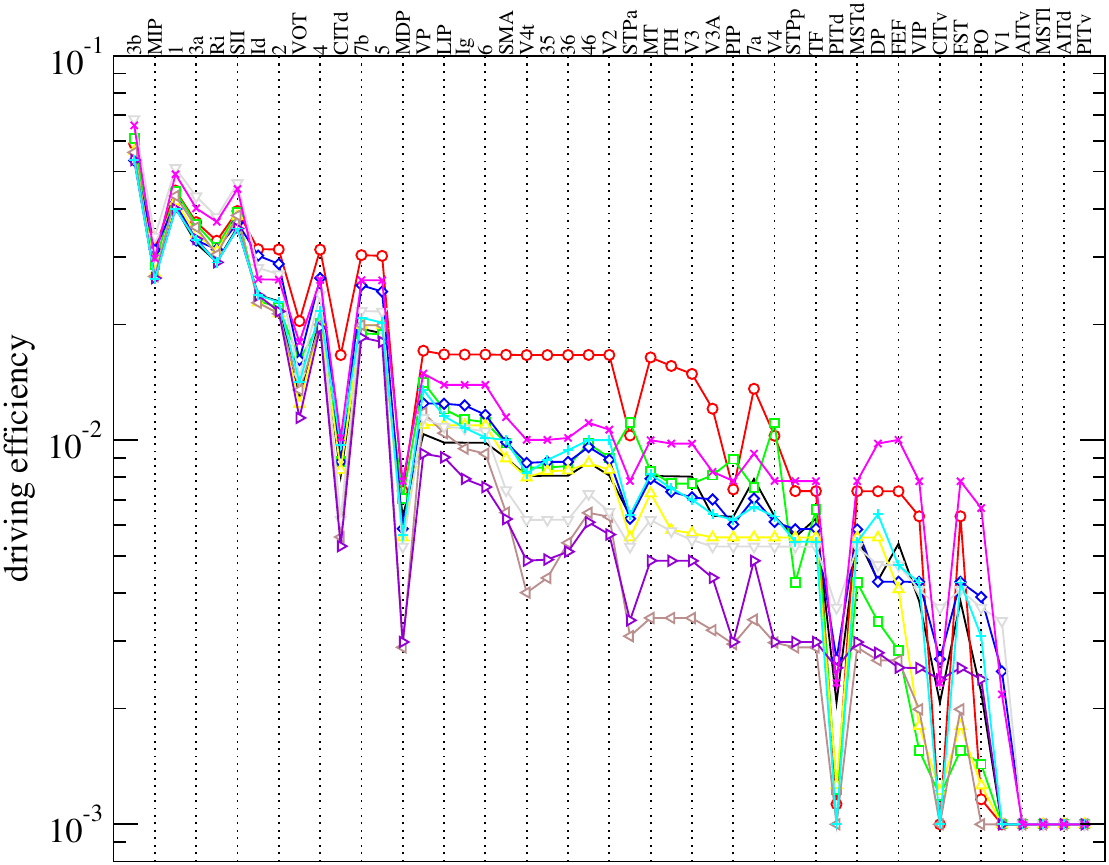}}
\caption{\label{fig:roess}
Driving efficiency of nodes with R\"{o}ssler oscillators on the Macaque cortex
network. Nodes are ordered from left to right by decreasing dynamical
influence. Symbols distinguish 10 independent random choices of initial
conditions of the nodes in the network and the driving node.
}
\end{figure*}

\subsection*{Driving coupled oscillators}

Further simulations of the driven system of phase-coupled oscillators are
performed, on randomly grown networks with (a) geometric (exponential) in- and
out-degree distributions and (b) a broad (scale-free) distribution of out-degree
and a geometric distribution of in-degree. The rank-order correlation between
driving efficiency and several centrality measures is shown in
Figure~\ref{fig:rhorscc} as a function of network size.

Next we replace the Kuramoto oscillators by R\"{o}ssler oscillators diffusively
coupled by all three variables. The autonomous dynamics evolves as
\begin{equation} \nonumber
\begin{array}{llll}
\dot{x}_i& = & -y_i-z_i      &+ \eta \sum_{j=1}^N K_{ij} (x_j-x_i)\\
\dot{y}_i& = & x_i + 0.2 y_i &+ \eta \sum_{j=1}^N K_{ij} (y_j-y_i)\\
\dot{z}_i& = & 0.2+z_i(x_i-9)&+ \eta \sum_{j=1}^N K_{ij} (z_j-z_i)\\
\end{array}~.
\end{equation}
with $(x_i,y_i,z_i)$ the three-dimensional state vector of node $i$ and the
coupling matrix $K$ of the network. The coupling strength is set $\eta=10$.
As the initial condition, all nodes in the
network obtain the same state $x_i(0) = x^{(0)}$, $y_i(0) = y^{(0)}$, $z_i(0) =
z^{(0)}$, $i\in\{1,\dots,N\}$ such that these oscillators are synchronous at the
beginning. We draw $x^{(0)}$, $y^{(0)}$, $z^{(0)}$ as independent standard random
numbers, i.e. from the uniform distribution on the unit interval. The driving
node is initialized independently, drawing standard random numbers for all three
variables. Then, as in the previous scenario with the Kuramoto oscillators,
a directed coupling (also of strength $\eta$) is established from the driving
node to a chosen node $i$. The time $T_i$ until synchronization is measured.
Synchronization is reached when for each node $j$ the absolute difference
between the driving node's state and the state of node $j$ is below $\epsilon$
in all three variables, with $\epsilon=10^{-4}$. If synchronization is not
reached by time $t=10^3$, $T_i=10^3$ is assigned. The driving efficiency of
node $i$ is the inverse of the time until synchronization when the driving
node feeds into node $i$ as described. The Macaque cortex network is used
for the simulations.

Predictive power of centrality measures is similar to the case of Kuramoto
oscillators. Driving efficiency has a rank order correlation of $0.957 \pm
0.010$ with dynamical influence, $0.646 \pm 0.012$ with degree ratio, $-0.013
\pm 0.0.011$ with shell index,  and $-0.072 \pm 0.007$ with betweenness. These
estimates are obtained as mean values and standard deviations over 10
independent realizations (initial conditions). Figure \ref{fig:roess} shows the
driving efficiencies obtained in these realizations, with nodes ordered by
dynamical influence.

\end{document}